\documentclass{PoS}
\usepackage{amsmath}
\usepackage{tikz}
\usetikzlibrary{calc,matrix,scopes,arrows}
\usepackage[textsize=small,shadow]{todonotes}
\usepackage{bbold}
\usepackage[utf8]{inputenc}
\usepackage[T1]{fontenc}
\tikzset{>=stealth'}

\title{Double Field Theory on Group Manifolds\\in a Nutshell}

\ShortTitle{Double Field Theory on Group Manifolds}

\author{Ralph Blumenhagen%
\\
      Max-Planck-Institute for Physics, Munich\\
      E-mail: \email{blumenha@mpp.mpg.de}}

\author{\speaker{Pascal du Bosque}%
\\
      Max-Planck-Institute for Physics, Munich\\
      Arnold-Sommerfeld-Center for Theoretical Physics, Munich \\
      E-mail: \email{dubosque@mpp.mpg.de}}
      
\author{Falk Hassler%
\\
      University of North Carolina, Chapel Hill\\
      E-mail: \email{fhassler@unc.edu}}
      
\author{Dieter L\"ust%
\\
      Max-Planck-Institute for Physics, Munich\\
      Arnold-Sommerfeld-Center for Theoretical Physics, Munich \\
      E-mail: \email{luest@mpp.mpg.de}}      

\abstract{We give a brief overview of the current status of Double Field Theory on Group Manifolds (DFT$_\mathrm{WZW}$). Therefore, we start by reviewing some basic notions known from Double Field Theory (DFT) and show how they extend/generalize into the framework of Double Field Theory on Group Manifolds. In this context, we discuss the relationship between both theories and the transition from DFT$_\mathrm{WZW}$ to DFT. Furthermore, we address some open questions and present an outlook into our current research.}

\FullConference{Corfu Summer Institute 2016 "School and Workshops on Elementary Particle Physics and Gravity"\\
		 31 August - 23 September, 2016\\
		 Corfu, Greece}

\begin{document}

\section{Intoduction}
Dualities are an important part of string theory. S- and T-Duality connect the five different perturbative string theories in a fascinating non-trivial way. Studying them led to a much deeper understanding of non-perturbative objects in string theory and ultimately culminated in the uncovering of M-Theory. However, as these theories are very complicated, only their low-energy effective actions are accessible in general. At this level, T-Duality is not manifest anymore in the NS/NS sector of $N=2$ Supergravity (SUGRA).

This motivated the development of Double Field Theory (DFT) \cite{Tseytlin:1990va, Siegel:1993th, Hull:2009mi, Hull:2009zb, Hohm:2010pp, Hohm:2010jy,  Hohm:2011ex, Aldazabal:2013sca, Hohm:2013bwa}. DFT makes the T-Duality group O($d,d,\mathbb{Z}$) of a $d$-dimensional torus manifest by introducing a doubled space \cite{Tseytlin:1990va, Siegel:1993th, Tseytlin:1990nb, Dabholkar:2002sy, Dabholkar:2005ve, Hull:2006va} which consists of the usual coordinates dual to momentum modes and additional coordinates dual to winding modes. The Buscher rules \cite{Buscher:1987sk} are implemented as global O($d,d$) transformations acting on the generalized metric. By reformulating SUGRA in the DFT framework, these transformations have a very natural interpretation as an exchange of winding and momentum modes. It should be kept in mind that the notion of the $2d$ doubled coordinates is totally equivalent to implementing $d$ left-moving and $d$ right-moving coordinates on the closed string.

However, DFT possesses some conceptual subtleties. These issues mainly arise as a consequence of the strong constraint, required for the consistent low-energy description of the theory. The strong constraint is a remainder of the toroidal background used in the derivation of DFT \cite{Hull:2009mi}. It prohibits winding and momentum excitations in the same direction. If we violate the strong constraint, we could not choose a torus radius in the each direction to make all fields in the effective theory lighter than the string scale. This is a severe issue, as either momentum or winding modes become heavier than the first massive string excitation and thus spoil an interpretation as a low-energy effective action. On the other hand, employing the strong constraint restricts DFT to the well-known NS/NS sector of SUGRA. Hence, it results in a rewriting of SUGRA. Nevertheless, it provides a striking tool in understanding these scenarios. Another questions regards the relaxation of the strong constraint, and the occurrence of so-called non-geometric fluxes \cite{Dabholkar:2005ve, Hull:2006va,Hull:2004in,Hull:2009sg, Andriot:2012wx} and whether they can be consistently included into the DFT framework. These are inspired by generalized Scherk-Schwarz compactifications that give rise to gauged SUGRAs inaccessible by flux compactifications from the SUGRA regime \cite{Dabholkar:2002sy, Hull:2005hk,Aldazabal:2011nj,Grana:2012rr,Aldazabal:2013mya,Berman:2013cli,Hassler:2014sba}. Some of these theories can be uplifted to string theory, but in general their status is unknown \cite{Dabholkar:2002sy,Hassler:2014sba, Condeescu:2012sp, Condeescu:2013yma}.

In order to resolve these issues, an alternative theory with doubled coordinate space called Double Field Theory on Group Manifolds (DFT$_\mathrm{WZW}$) has been proposed \cite{Blumenhagen:2014gva, Blumenhagen:2015zma, Bosque:2015jda, Hassler:2016srl}. Starting from Closed String Field Theory (CSFT) calculations at tree level up to cubic order in fields and leading order of $\alpha'$, an action and the according gauge transformations have been derived from a Wess-Zumino-Witten (WZW) model and its corresponding Ka$\check{\text{c}}$-Moody algebra \cite{Blumenhagen:2014gva}. A related approach can be found in \cite{Hohm:2015ugy}. In this context, the doubling of the coordinates refers to the left- and right-moving currents of the WZW model on a group manifold. Intriguingly, this theory does not exactly reproduce all results known from toroidal DFT. Instead, the action, the gauge transformations and the strong constraint receive corrections from the non-trivial string background \cite{Blumenhagen:2015zma}. While the fluctuations still have to fulfill the strong constraint, the background fields only have to satisfy the much weaker Jacobi-Identity. Subsequently, we are left with a consistent tree-level description of non-geometric backgrounds. Furthermore, the relaxation of the strong constraint for the background fields is closely related to the closure constraint in the original flux formulation \cite{Aldazabal:2013sca, Grana:2012rr, Geissbuhler:2011mx, Hohm:2013nja, Geissbuhler:2013uka} and in turn allows for truly non-geometric backgrounds in toroidal DFT, which are inaccessible from any geometric configuration. All of these observations suggest that DFT$_\mathrm{WZW}$ can be considered as a generalization of DFT. It can be reformulated in form of a generalized metric formulation \cite{Blumenhagen:2015zma}, thus extrapolating the theory to all orders in fields, and in form of a flux formulation \cite{Bosque:2015jda}. Through the consequent use of covariant derivatives in this theory, it possesses an additional $2D$-diffeomorphism invariance as a result of the rigorous splitting between background fields and fluctuations in DFT$_\mathrm{WZW}$ \cite{Blumenhagen:2015zma}. In this paper we want to present a short summary of DFT$_\mathrm{WZW}$ and its salient features.

First, we briefly review the traditional formulation of DFT in section \ref{sec:dft}. Afterwards, we want to turn to DFT$_\mathrm{WZW}$, in the next chapter \ref{sec:dftwzw}, while trying to highlight the similarities and differences of both theories. In the conclusion \ref{sec:conclusion}, we will address open questions and give an update about our current research.

\section{Double Field Theory}\label{sec:dft}
\subsection{Action}
In its most basic form, DFT captures the NS/NS section of $N=2$ Supergravity which arises as a low energy effective theory for massless string excitations in $D$-dimensions. Its action
\begin{equation}\label{eqn:S_SUGRA}
S_{NS} = \int d^D x \sqrt{-g} \, e^{-2\phi} \left[ R + 4 (\partial \phi)^2 - \frac{1}{12} H_{ijk} H^{ijk} \right]
\end{equation}
involves the metric $g_{ij}$, the 2-form field $b_{ij}$, and the dilaton $\phi$. SUGRA combines Supersymmetry (SUSY) with general relativity (GR) by writing an action which is invariant under local SUSY transformations. It is the low energy limit ($E \ll m_s$) of Superstring theory. Here, we are only interested in the bosonic part of the action. Hence, we do not consider SUSY transformations which exchange bosons and fermions.

In DFT, the SUGRA action gets geometrized. Meaning, it is expressed in terms of a generalized Ricci scalar $\mathcal{R}$ \cite{Hull:2009mi,Hohm:2010pp, Hohm:2010jy} and a density involving the generalized dilaton $d$. The corresponding action on the doubled space now looks very similar to the Einstein-Hilbert action of general relativity. Specifically, it takes the form
\begin{equation}
S_{DFT} = \int d^{2D} X \, e^{-2d} \mathcal{R}\left( \mathcal{H}, d \right)\,.
\end{equation}
Indeed, this action captures the same dynamics of \eqref{eqn:S_SUGRA}, if one introduces the generalized curvature scalar
\begin{align}
\mathcal{R} \equiv \;\; &4\mathcal{H}^{MN} \partial_M d\, \partial_N d - \partial_M \partial_N \mathcal{H}^{MN} -4\mathcal{H}^{MN} \partial_M d\, \partial_N d \\ +& 4 \partial_M \mathcal{H}^{MN} \partial_N d +\frac{1}{8} \mathcal{H}^{MN} \partial_M \mathcal{H}^{KL} \partial_{N} \mathcal{H}_{KL} - \frac{1}{2} \mathcal{H}^{MN} \partial_N \mathcal{H}^{KL} \partial_{L} \mathcal{H}_{MK} \nonumber
\end{align}
in $2D$-dimensions. We are going to discuss under which transformations $\mathcal{R}$ can be understood as a scalar in a moment. Let us first turn to the notion and symbols, we use in the expression which we have just presented. The idea of a doubled space was first introduced by Tseytlin and Siegel \cite{Tseytlin:1990va,Siegel:1993th}. It comes with the coordinates
\begin{equation}
X^M = ( \tilde{x}_i, x^i ) 
\end{equation}
which consist of the physical coordinates $x^i$ and additional coordinates $\tilde{x}_i$, also called winding coordinates. Accordingly, we can also write down the doubled, partial derivative
\begin{equation}
  \partial_M = ( \tilde{\partial}^i, \partial_i ) \quad \text{with} \quad \partial_M X^N = \delta_M^N \,,
\end{equation}
which appears in the generalized curvature scalar. The indices are raised and lowered with the O($D,D$) invariant metric
\begin{equation}
  \eta^{MN}=\begin{pmatrix}
    0 & \delta_i^j \\
    \delta^i_j & 0
  \end{pmatrix}
\end{equation}
and its inverse $\eta_{MN}$. Now, we want to turn to the remaining two quantities occurring in the generalized Ricci scalar, the generalized metric $\mathcal{H}_{MN}(X)$ and the generalized dilaton $d(X)$. Both depend on the coordinates of the doubled space \cite{Hohm:2010pp} with the restriction that they have to fulfill the strong constraint we discuss in the next subsection. The generalized metric is given by
\begin{equation}
  \mathcal{H}_{MN}(X) = \begin{pmatrix}
    g^{ij} & -g^{ik} B_{kj} \\ B_{ik} g^{kj} & g_{ij} - B_{ik} g^{kl} B_{lj}
  \end{pmatrix}\,.
\end{equation}
It represents an O($D,D$) valued, symmetric matrix. The Buscher rules for abelian isometries in SUGRA can be rewritten in terms of the transformation
\begin{equation}
  \mathcal{H}_{MN} \rightarrow O^L{}_M \mathcal{H}_{LK} O^K{}_N \quad \text{and} \quad
  d \rightarrow d
\end{equation}
with
\begin{equation}
  O^M{}_N = \begin{pmatrix}
    m^i_j & \delta^{ij} - m^{ij} \\
    \delta_{ij} - m_{ij} & m^j_i
  \end{pmatrix} \quad\text{and}\quad
  m^i_j = m_{ij} = m^{ij} = \begin{cases} 1 & i=j=a \\
    0 & \text{otherwise}
  \end{cases}
\end{equation}
acting on the generalized metric. Thus, T-Duality for abelian isometries is a manifest global symmetry of the DFT action. Therefore, it allows to describe all the dual backgrounds in a unified way. Additionally, we identify the dilaton density by
\begin{equation}
e^{-2d} = \sqrt{-g} \, e^{-2\phi}\,.
\end{equation}

Furthermore, we want to raise the question whether it is possible to retrieve the SUGRA action from the DFT action. In fact, this is possible by setting the winding derivatives in the DFT action to zero
\begin{equation}
S_{DFT}  \quad \xrightarrow{\tilde{\partial}^i =  \,0 \,} \quad S_{NS}\,.
\end{equation}
Doing so, we recover the SUGRA action \eqref{eqn:S_SUGRA}. This restricts the fields to live on a $D$-dimensional subspace of the $2D$-dimensional doubled space-time.

\subsection{Gauge transformations}
Let us now turn to the generalized diffeomorphisms, under which the generalized Ricci scalar transforms as a scalar \cite{Hull:2009mi, Hull:2009zb, Hohm:2010pp}. Their infinitesimal version is mediated by the generalized Lie derivative
\begin{equation}
  \mathcal{\hat{L}}_{\xi} \mathcal{H}^{MN} = \xi^P \partial_P \mathcal{H}^{MN} + \big(\partial^M \xi_P - \partial_P \xi^M \big) \mathcal{H}^{PN} + \big(\partial^N \xi_P - \partial_P \xi^N \big) \mathcal{H}^{MP}\,.
\end{equation}
Applied on the generalized metric, it encodes the gauge transformations of the metric $g_{ij}$ and the 2-form $B_{ij}$. For the generalized dilaton, a density under generalized diffeomorphisms, we have
  \begin{equation}
   \mathcal{\hat{L}}_{\xi} d = - \frac{1}{2} \partial_M \xi^M + \xi^M \partial_M d\,, \quad\text{and}\quad    \mathcal{\hat{L}}_{\xi} e^{-2d} = \partial_M ( \xi^M e^{-2d} )\,.
  \end{equation}
Through its particular form, the generalized Lie derivative preserves the O$(D,D)$ structure of the theory. A good way to see this, is to apply it to the O($D,D$) metric invariant metric which yields
\begin{equation}
\mathcal{\hat{L}}_{\xi} \eta^{MN} = 0\,.
\end{equation}
Finally, we can check how the DFT action transforms under generalized diffeomorphisms. As expected for a symmetry, it is invariant as long as we impose the strong constraint \cite{Hull:2009mi}
\begin{equation}
\partial^M \partial_M ( A \cdot B ) = 0 \quad \forall\;\text{fields}\;A\,,B\,.
\end{equation}
It is a consequence of the level-matching condition during string scattering processes which are globally O($D,D$) covariant. This constraint restricts the fields of the theory considerably and allows for the construction of toroidal DFT to all orders in fields.  A canonical way to solve this condition is the SUGRA limit
\begin{equation}
\partial_M = \begin{cases} \partial_i\quad \text{if}\; M=i \\ 0\quad \;\text{else} \end{cases}
\end{equation}
which results in the action \eqref{eqn:S_SUGRA}. Last but not least, we want to take a closer look at the gauge algebra induced by the generalized Lie derivative. It closes involving the C-bracket, an O($D,D$) covariant extension of the Courant-bracket \cite{Hull:2009zb, Hohm:2010pp},
\begin{equation}
\big[  \mathcal{\hat{L}}_{\xi_1},  \mathcal{\hat{L}}_{\xi_2} \big] =  \mathcal{\hat{L}}_{[\xi_1, \xi_2]_C} \end{equation}
modulo terms that vanish under the strong constraint. The C-bracket is defined as
\begin{equation}
\big[ \xi_1, \xi_2 \big]_C^M = \xi_1^N \partial_N \xi_2^M - \xi_2^N \partial_N \xi_1^M - \frac{1}{2} \xi_{1N} \partial^M \xi_2^N + \frac{1}{2} \xi_{2N} \partial^M \xi_1^N \,.
\end{equation}
In contrast to the Lie bracket, the Jacobiator of the C-bracket does not vanish. Hence, the C-bracket generally does not generate a Lie algebra.

\subsection{Flux formulation}
An entirely equivalent way to express the DFT action is in the flux formulation \cite{Aldazabal:2013sca, Geissbuhler:2011mx, Hohm:2013nja, Geissbuhler:2013uka}. In order to recast the theory into this formulation, we need to decompose the generalized metric and O($D,D$) metric using a vielbein $E_A{}^I$$\in$ O($D,D$) by
\begin{equation}\label{eqn:decompgenmetric}
\mathcal{H}_{MN} = E^A{}_M \, S_{AB} \, E^B{}_N \quad \text{and} \quad \eta_{MN} = E^A{}_M \, \eta_{AB} \, E^B{}_N \,.
\end{equation}
This allows us to define the covariant fluxes
\begin{align}
\mathcal{F}_{ABC} &= \mathcal{\hat{L}}_{E_A} E_B{}^M E_{CM} = 3 \tilde{\Omega}_{[ABC]}\,, \\
\mathcal{F}_A &= - e^{2d} \mathcal{\hat{L}}_{E_A} e^{-2d} = 2 E_A{}^M \partial_M d + \tilde{\Omega}^B{}_{BA}\,,
\end{align}
which are scalars under O($D,D$) transformations. A quantity which frequently appears in this formulation is the Weitzenb\"ock connection (also called coefficients of anholonomy)
\begin{equation}
  \tilde{\Omega}_{ABC} = E_A{}^I \partial_I E_B{}^J E_{CJ}\,.
\end{equation}
Plugging the decomposition ansatz \eqref{eqn:decompgenmetric} into the DFT action, we obtain
\begin{align}
S= \int d^{2D}X \, e^{-2d} ( &\mathcal{F}_A \mathcal{F}_B S^{AB} + \frac{1}{4} \mathcal{F}_{ACD} \mathcal{F}_B{}^{CD} S^{AB} \\ - \frac{1}{12} &\mathcal{F}_{ACE} \mathcal{F}_{BDF} S^{AB} S^{CD} S^{EF}  )\,.
\end{align}
This action is manifestly invariant under generalized diffeomorphisms. However, now there are also double Lorentz transformations which act on the flat indices $A, B, C, \dots$\,. Even if $\mathcal{F}_{ABC}$ and $\mathcal{F}_A$ do not transform covariantly under this symmetry, the complete action is invariant under it \cite{Geissbuhler:2013uka}. Moreover, using an appropriate compactification ansatz for the fluxes, we recover the bosonic sector of half-maximal, electrically gauged SUGRA \cite{Geissbuhler:2011mx, Hohm:2013nja}. The equations of motion \cite{Aldazabal:2013sca,Geissbuhler:2013uka} can be derived by varying the action with respect to the generalized vielbein $E_A{}^I$ and the dilaton $d$ respectively
\begin{equation}
\mathcal{G} = -2 \mathcal{R} = 0 \quad \text{and} \quad \mathcal{G}^{[AB]} = 0
\end{equation}
with
\begin{equation}
\mathcal{G}^{AB} = 2 \left( S^{DA} - \eta^{DA} \right) \partial^B \mathcal{F}_D + \left( \mathcal{F}_D - \partial_D \right) \check{\mathcal{F}}^{DAB} + \check{\mathcal{F}}^{CDA} \mathcal{F}_{CD}{}^B
\end{equation}
and
\begin{equation}
\check{\mathcal{F}}^{ABC} = \frac{3}{2} \mathcal{F}_D{}^{BC} S^{AD} - \frac{1}{2} \mathcal{F}_{DEF} S^{AD} S^{BE} S^{CF} - \mathcal{F}^{ABC}\,.
\end{equation}
Again, for the strong constraint solution $\tilde{\partial}^i = 0$ everything reduces to SUGRA as one would expect.

\section{Double Field Theory on Group Manifolds}\label{sec:dftwzw}
\subsection{Basics}
Starting from a group manifold, instead of a torus, it is possible to obtain a doubled theory as well. It is called Double Field Theory on Group Manifolds (DFT$_\mathrm{WZW}$) \cite{Blumenhagen:2014gva} and was derived from a Wess-Zumino-Witten (WZW) model on the worldsheet. The doubling of the coordinates is related to the left- and right-moving Ka$\check{\text{c}}$-Moody current algebras of the WZW model. Primary fields of the CFT are represented as scalar functions on the group manifold $G=G_\mathrm{L}\times G_\mathrm{R}$. Its left-moving (chiral) part is equipped with the coordinates $x^i$ and the flat derivatives $D_a = e_a{}^i \partial_i$. In the definition of the flat derivatives, we use the background vielbein $e_a{}^i $ on $G_\mathrm{L}$ such that they implement the algebra
\begin{equation}\label{eqn:comm1}
\left[ D_a, D_b \right] = F_{ab}{}^c D_c\,.
\end{equation}
The background vielbein originates from a Maurer-Cartan form through the relation
\begin{equation}
\omega_\gamma = t_a \, e^a{}_i \, dx^i\,,\quad \text{where} \quad e^a{}_i = \mathcal{K}\left( t^a, \gamma^{-1} \partial_i \gamma \right) \,.
\end{equation}
Here, $t_a$ denotes the generators of the Lie algebra $\mathfrak{g}_\mathrm{L}$ where indices are raised and lowered with the flat metric $\eta_{ab}$ given by the Killing form\footnote{For simplicity we assume that $G$ is semisimple. However, the equations we discuss later also hold for a more general case.} $\mathcal{K}$ as $\eta_{ab}=\mathcal{K}(t_a, t_b)$. The flat derivatives act on patches of the group manifold, as opposed to the generators $t_a$ which act on an abstract notion of the related vector space. Therefore, the functions on these patches, on which the flat derivatives are acting, are a representation of the universal enveloping algebra of the associated Lie algebra.

The same holds for the right-moving (anti-chiral) part with the coordinates $\bar{x}^i$ and the flat derivative $D_{\bar{a}} = e_{\bar{a}}{}^{\bar{i}} \partial_{\bar{i}}$ which yields the following algebra
\begin{equation}\label{eqn:comm2}
\left[ D_{\bar{a}}, D_{\bar{b}} \right] = F_{\bar{a}\bar{b}}{}^{\bar{c}} D_{\bar{c}}\,.
\end{equation}
In this context, we treat the left-movers and right-movers independently of each other. Further notice that the unimodularity of the Lie algebras allow for integration by parts on the group manifold. 

At this point, we can perform tree-level Closed String Field Theory (CSFT) calculation up to cubic order in the fields and leading order in $\alpha'$ to derive the action and gauge transformations of the theory. To this end it is necessary to obtain the correlation functions of Kac-Moody primary fields. They are explicitly given in \cite{Blumenhagen:2014gva}. It should be mentioned that the chiral and anti-chiral currents posses the same underlying Kac-Moody algebra. Subsequently, we find the following action
\begin{equation}\label{eqn:DFTwzwaction}
{\begin{aligned}
  (2 \kappa^2)S &= \int d^{2D} X\sqrt{|H|}\, \Big[ 
      \epsilon_{a\bar b} \,\square \epsilon^{a\bar b} + (D^{\bar b} \epsilon_{a\bar b})^2
      +  (D^a \epsilon_{a \bar b})^2 + 4 \tilde d\, D^a D^{\bar b} \epsilon_{a\bar b}  \\
      &- 4 \tilde d\, \square \tilde d
    -2 \epsilon_{a\bar b} \bigl( D^a \epsilon_{c\bar d}\, D^{\bar b} \epsilon^{c\bar d} - 
      D^a \epsilon_{c \bar d}\, D^{\bar d} \epsilon^{c\bar b} - D^c \epsilon^{a\bar d}\, D^{\bar b} 
      \epsilon_{c\bar d} \bigr) \\
    & +2 \epsilon_{a\bar b} \bigl( F^{ac}{}_d\, D^{\bar e} \epsilon^{d\bar b} \;\epsilon_{c\bar e} 
      + F^{\bar b\bar c}{}_{\bar d}\, D^e \epsilon^{a\bar d}\; \epsilon_{e\bar c} \bigr)
      + \frac{2}{3} F^{ace}\, F^{\bar b\bar d\bar f}\, \epsilon_{a\bar b}\, \epsilon_{c\bar d}\,
      \epsilon_{e\bar f} \\
    & +\tilde d \bigl( 2 (D^a \epsilon_{a\bar b})^2 + 2 (D^{\bar b} \epsilon_{a\bar b})^2 + 
      (D_c \epsilon_{a\bar b})^2 + (D_{\bar c} \epsilon_{a\bar b})^2  \\ &+ 4 \epsilon^{a\bar b}
      ( D_a D^c \epsilon_{c\bar b} + D_{\bar b} D^{\bar c} \epsilon_{a\bar c} )
      \bigr)
    -8 \epsilon_{a\bar b}\, \tilde d\, D^a D^{\bar b} \tilde d + 4 {\tilde d}^2\, \square \tilde d \Big]
  \end{aligned}}
\end{equation}
and the corresponding gauge transformations
\begin{equation}
{\begin{aligned}
  \delta_{\lambda} \epsilon^{a\bar b} =& - D^{\bar b} \lambda^a + 
    D^a \lambda_c \epsilon^{c\bar b} - D_c \lambda^a \epsilon^{c\bar b} + \lambda^c D_c \epsilon^{a \bar b}
    + F^a{}_{cd}\, \lambda^c \epsilon^{d\bar b}  \\ 
  & -D^a \lambda^{\bar b} + D^{\bar b} \lambda_{\bar c} \epsilon^{a\bar c} - 
    D_{\bar c} \lambda^{\bar b} \epsilon^{a \bar c} + \lambda^{\bar c} D_{\bar c} \epsilon^{a \bar b}
    + F^{\bar b}{}_{\bar c \bar d} \lambda^{\bar c}\, \epsilon^{a \bar d} \\
    \delta_\lambda \tilde d =& -\frac{1}{2} D_a \lambda^a + \lambda_a\, D^a \tilde d - 
    \frac{1}{2} D_{\bar a} \lambda^{\bar a} + \lambda_{\bar a}\,
    D^{\bar a} {\tilde d} \label{eqn:gaugetrafos}
  \end{aligned}}
\end{equation}
in terms of the fluctuation fields $\epsilon^{a\bar{b}}$ of the metric, and the $B$-field, as well as the fluctuation dilaton $\tilde{d}$.

\subsection{Gauge transformations}
All DFT${}_\mathrm{WZW}$ results discussed so far use unbarred coordinates $x^a$ and barred coordinates $x^{\tilde{a}}$ with the corresponding flat derivatives.  Unbarred flat derivatives $D_a$ only act on unbarred coordinates $x^a$ and vice versa. To make the structure of the theory manifest (and at the same time simplify all expressions considerably), we combine them into the doubled quantities
\begin{equation}
x^A = \begin{pmatrix} x^a \\ x^{\bar a} \end{pmatrix}\quad \text{and the doubled flat derivative} \quad D_A = \begin{pmatrix} D_a \\ D_{\bar a} \end{pmatrix}\,.
\end{equation}
Additionally, we define a doubled vielbein by
\begin{equation}
E_A{}^I = \begin{pmatrix}
e_a{}^i & 0 \\ 0 & e_{\bar a}{}^{\bar i }
\end{pmatrix}
\end{equation}
which yields the commutation relations \eqref{eqn:comm1} and \eqref{eqn:comm2} on the doubled space
\begin{equation}
\left[ D_A, D_B \right] = F_{AB}{}^C \, D_C
\end{equation}
with the structure coefficients given by
\begin{equation}
{F_{AB}}^C = 2 \Omega_{[AB]C} = 
\begin{cases} {F_{ab}}^c \\ {F_{\bar{a}\bar{b}}}^{\bar{c}} \\ 0\quad\text{otherwise\,.} \end{cases}\
\end{equation}
Now, we can rewrite the gauge transformations as
\begin{equation}
  \mathcal{L}_{\xi} \mathcal{H}^{AB} = \xi^C \nabla_C \mathcal{H}^{AB} + \big(\nabla^A   \xi_C - \nabla_C \xi^A \big) \mathcal{H}^{CB} + \big(\nabla^B \xi_C - \nabla_C \xi^B \big) \mathcal{H}^{AC}\,.
  \end{equation} 
Here, we introduced the covariant derivative
\begin{equation}
\nabla_A V^B = D_A V^B + \frac{1}{3} {F^B}_{AC} V^C
\end{equation}
and expanded the fluctuations $\epsilon^{a\bar{b}}$ by
\begin{equation}\label{eqn:expgenmetric}
\mathcal{H}^{AB} = \text{exp}(\epsilon^{AB}) = S^{AB} + \epsilon^{AB} + \frac{1}{2} \epsilon^{AC} S_{CD} \epsilon^{DB} + \ldots
\end{equation}
with
\begin{equation}
  S^{AB} = \frac12 \begin{pmatrix} \eta^{ab} & 0 \\
    0 & \eta^{\bar a\bar b}
  \end{pmatrix} \quad \text{and} \quad 
  \epsilon^{AB} = \begin{pmatrix} 0 & \epsilon^{a\bar{b}} \\  \epsilon^{\bar{a}b} & 0 \end{pmatrix}\,.
\end{equation}
The striking difference to original DFT lies in the occurrence of terms containing structure coefficients from the background, which cannot be seen in DFT due to is toroidal structure, as they vanish in this limit.

Successively, we obtain for the fluctuation dilaton, in the second part of \eqref{eqn:gaugetrafos},
\begin{equation}
  \delta_\xi \tilde{d} = \mathcal{L}_\xi \tilde{d} = \xi^A D_A \tilde{d} - \frac{1}{2} D_A \xi^A
    \quad\text{with}\quad
  \nabla_A d = D_A \tilde{d}\,.
\end{equation}
As can be checked easily, this new generalized Lie derivative still preserves the O($D,D$) structure
\begin{equation}
  \mathcal{L}_\xi \eta^{AB} = 0 \quad \text{with} \quad
  \eta^{AB} = \frac12 \begin{pmatrix} \eta^{ab} & 0 \\ 0 & - \eta^{\bar a\bar b} \end{pmatrix} \,.
\end{equation}
At this point, we want to turn to the closure of the gauge algebra and it is therefore vital to study the relation
\begin{equation}
\big[  \mathcal{\hat{L}}_{\xi_1},  \mathcal{\hat{L}}_{\xi_2} \big] =  \mathcal{\hat{L}}_{[\xi_1, \xi_2]_C} \; \text{modulo strong and closure constraint}
\end{equation}
which is governed by the C-bracket
\begin{equation}
\big[ \xi_1, \xi_2 \big]_C^A = \xi_1^B \nabla_B \xi_2^A - \xi_2^B \nabla_B \xi_1^A - \frac{1}{2} \xi_{1B} \nabla^A \xi_2^B + \frac{1}{2} \xi_{2B} \nabla^A \xi_1^B\,.
\end{equation}
Again, it is straightforward to see that in the toroidal limit this gives rise to the C-bracket of original DFT. A similar structure for the generalized Lie derivative is known from the flux formulation \cite{Geissbuhler:2013uka}. The strong constraint in this context is given by
\begin{equation}
  D^A D_A \, ( f \cdot g ) = 0 \quad \forall \; \text{fluctuations}\,f,\,g\,,
\end{equation}
as fields which are changed by gauge transformations should still satisfy the level matching condition. For transformations of cubic order this is in general not the case. Therefore, it is necessary that all fields have to be projected into the kernel of the level matching operator $\Delta$. However, in order to avoid always having to perform this explicit projection, we impose the strong constraint. It makes sure that the string product is always level matched. In the DFT$_\mathrm{WZW}$ setup it is generally not trivial to solve this constraint. A detailed procedure to do so is presented in \cite{Hassler:2016srl}.

Moreover, the closure constraint for the background (Jacobi equation) reads
\begin{equation}
F_{E[AB} {F^E}_{C]D} = 0\quad\text{(background fields)}\,.
\end{equation}
On a group manifold, it is always fulfilled.

Hence, the strong constraint of DFT$_{\mathrm{WZW}}$ gets corrected by the background vielbein $E_A{}^I$ when comparing it to toroidal DFT and the gauge transformations are extended by additional terms containing structure coefficients. Furthermore, we see the appearance of another constraint for the background fields due to the underlying group manifold. This term vanishes in the toroidal (DFT) limit.

\subsection{Action}
From the CSFT framework its already clear that DFT$_{\mathrm{WZW}}$ should be invariant under its gauge transformations up to cubic order. Nevertheless, an explicit verification was performed in \cite{Blumenhagen:2015zma}. It provides an additional consistency check and extends the CSFT results from cubic order to arbitrary order in the fields.

Now, we recast the DFT$_{\mathrm{WZW}}$ action \eqref{eqn:DFTwzwaction} using the generalized metric $\mathcal{H}^{AB}$ introduced in \eqref{eqn:expgenmetric}. The result is \cite{Blumenhagen:2015zma}
\begin{equation}
S_{DFT{_\mathrm{WZW}}} = \int d^{2D}X \; e^{-2d} \mathcal{R}(\mathcal{H}, d)
\end{equation}
with the generalized curvature scalar
\begin{align}
\label{eqn:genriccidftwzw}
\mathcal{R} &\equiv \,\,4\mathcal{H}^{AB} \nabla_A d\, \nabla_B d - \nabla_A \nabla_B \mathcal{H}^{AB} -4\mathcal{H}^{AB} \nabla_A d\, \nabla_B d \\ &\,\,\,+ 4 \nabla_A \mathcal{H}^{AB} \nabla_B d +\frac{1}{8} \mathcal{H}^{AB} \nabla_A \mathcal{H}^{CD} \nabla_{B} \mathcal{H}_{CD} - \frac{1}{2} \mathcal{H}^{AB} \nabla_B \mathcal{H}^{CD} \nabla_{D} \mathcal{H}_{AC} \nonumber \\ &{\;\;\,+\,\frac{1}{6} F_{ACE} F_{BDF} \mathcal{H}^{AB} S^{CD} S^{EF}}\,. \nonumber
\end{align}
This action is invariant under generalized diffeomorphisms when the strong constraint and closure constraint are imposed. It should be stressed again that this result holds to all order in fields in the generalized metric formulation, and not just up to cubic order. Furthermore, the last term of the generalized Ricci scalar \eqref{eqn:genriccidftwzw} is crucial for the invariance under generalized diffeomorphisms and cannot be seen in toroidal DFT, since the structure coefficients vanish in this limit. Moreover, due to the geometric structure of the theory it possesses an $2D$-diffeomorphism invariance (standard Lie derivative) in contrast to toroidal DFT.

\subsection{Flux formulation}
Before recasting the theory in its flux formulation, we have to fix the building blocks of this formulation, the covariant fluxes. Analogous to toroidal DFT, we can decompose the generalized metric and the flat O($D,D$) invariant metric using a fluctuation vielbein $\tilde{E}_{\hat{A}}{}^B$ $\in$ O($D,D$) according to
\begin{equation}
  \mathcal{H}_{AB} = \tilde{E}^{\hat{C}}{}_A \, S_{\hat{C}\hat{D}} \, \tilde{E}^{\hat{D}}{}_B \quad \text{and} \quad \eta_{AB} = \tilde{E}^{\hat{C}}{}_A \, \eta_{\hat{C}\hat{D}} \, \tilde{E}^{\hat{D}}{}_B
\end{equation}
We then combine it with the background vielbein $E_A{}^I$ $\in$ GL($2D$) to form a so-called composite vielbein
\begin{equation}
  \mathcal{E}_{\hat{A}}{}^I = \tilde{E}_{\hat{A}}{}^B E_B{}^I \,.
\end{equation}
The additional structure is best illustrated by the diagram
\begin{equation}\label{eqn:groupindices}
  \tikz[baseline]{\matrix (m) [ampersand replacement=\&, matrix of math nodes, column sep=4em] {\text{O}(1,D-1)\times\text{O}(D-1,1) \& \text{O}(D,D) \& \text{GL}(2 D) \\};
    \draw[->] (m-1-3) -- (m-1-2) node[midway, above] {$\eta_{IJ}$} node[midway, below] {$E_B{}^I$};
    \draw[->] (m-1-2) -- (m-1-1) node[midway, above] {$\mathcal{H}_{AB}$} node[midway, below] {${\tilde E}_{\hat A}{}^B$};}\,.
\end{equation}
Starting from a $2D$-dimensional smooth manifold $M$ equipped with a pseudo Riemannian metric $\eta$, with split signature, it reduces the manifold's structure group GL($2D$) to O($D,D$). Then, the corresponding frame bundle on $M$ is given through the background vielbein $E_A{}^I$. Subsequently, the generalized metric $\mathcal{H}_{AB}$ further reduces the structure group to $\text{O}(1,D-1)\times\text{O}(D-1,1)$, the double Lorentz group, which is represented by the fluctuation frame $\tilde{E}_{\hat{A}}{}^B$. Original DFT lacks the information contained in the background vielbein $E_B{}^I$ and therewith the right part of this diagram.

The covariant fluxes are now given by
\begin{equation}
\mathcal{F}_{\hat{A}\hat{B}\hat{C}} = \mathcal{L}_{\mathcal{E}_{\hat{A}}} \mathcal{E}_{\hat{B}}{}^M \mathcal{E}_{\hat{C}M} = 3 \tilde{\Omega}_{[\hat{A}\hat{B}\hat{C}]} \overbrace{\,+\, 2 \Omega_{[\hat{A}\hat{B}]\hat{C}}}^{\text{structure coefficients}}\,,
\end{equation}
\begin{equation}
\mathcal{F}_{\hat{A}} = - e^{2d} \mathcal{L}_{\mathcal{E}_{\hat{A}}} e^{-2d} = 2 \mathcal{E}_{\hat{A}}{}^M \partial_M d + \tilde{\Omega}^{\hat{B}}{}_{\hat{B}\hat{A}}\,.
\end{equation}
In contrast to DFT, the coarivant fluxes $\mathcal{F}_{\hat{A}\hat{B}\hat{C}}$ in DFT$_{\mathrm{WZW}}$ get an additional contribution from the structure coefficients of the underlying group manifold. Again, they are scalars under generalized diffeomorphisms by construction. Due to the geometric structure of DFT$_{\mathrm{WZW}}$ they possess an $2D$-diffeomorphism invariance as well. Recasting the DFT$_{\mathrm{WZW}}$ action yields
\begin{align}
  S= \int d^{2D}X \, e^{-2d} ( &\mathcal{F}_{\hat{A}} \mathcal{F}_{\hat{B}} S^{\hat{A}\hat{B}} + \frac{1}{4} \mathcal{F}_{\hat{A}\hat{C}\hat{D}} \mathcal{F}_{\hat{B}}{}^{\hat{C}\hat{D}} S^{\hat{A}\hat{B}} \\ - \frac{1}{12} &\mathcal{F}_{\hat{A}\hat{C}\hat{E}} \mathcal{F}_{\hat{B}\hat{D}\hat{F}} S^{\hat{A}\hat{B}} S^{\hat{C}\hat{D}} S^{\hat{E}\hat{F}}  ) \,. \nonumber
\end{align}
This action also has an invariance under double Lorentz transformations like its DFT counterpart. It should be mentioned that this action does not incorporate any strong constraint violating terms due to independently vanishing chiral and anti-chiral total central charges. An appropriate compactification ansatz leads again to half-maximal, electrically gauged SUGRA. The equations of motion are then given by variation after the fluctuation vielbein $\tilde{E}_{\hat{A}}{}^B$ and the dilaton $d$. We obtain
\begin{equation}
  \mathcal{G} = -2 \mathcal{R} = 0 \quad \text{and} \quad \mathcal{G}^{[\hat{A}\hat{B}]} = 0
\end{equation}
with
\begin{equation}
\mathcal{G}^{\hat A \hat B} = 2 S^{\hat D \hat A} D^{\hat B} \mathcal{F}_{ \hat D} + \left( \mathcal{F}_{ \hat D} - D_{ \hat D} \right) \check{\mathcal{F}}^{\hat D \hat A \hat B} + \check{\mathcal{F}}^{\hat C \hat D \hat A} \mathcal{F}_{\hat C \hat D}{}^{ \hat B}
\end{equation}
and
\begin{equation}
\check{\mathcal{F}}^{ABC} = \left( - \frac{1}{2} S^{\hat A \hat B} S^{\hat C \hat D} S^{\hat E \hat F} + \frac{1}{2} S^{\hat A \hat B} \eta^{\hat C \hat D} \eta^{\hat E \hat F} + \frac{1}{2} \eta^{\hat A \hat B} S^{\hat C \hat D} \eta^{\hat E \hat F} + \frac{1}{2} \eta^{\hat A \hat B} \eta^{\hat C \hat D} S^{\hat E \hat F} \right) \mathcal{F}_{\hat D \hat E \hat F}\,.
\end{equation}

\subsection{Transition to toroidal DFT}
In order to make contact with the traditional formulation of DFT, we need to impose an extra constraint \cite{Blumenhagen:2015zma, Bosque:2015jda}. Generally, there exist two possible ways to do this. One possibility is to set the background vielbein $E_A{}^I = \text{constant}$ which leads to vanishing structure coefficients and thus yields toroidal DFT. The other, a more general and involved procedure requires to employ the so-called extended strong constraint
\begin{equation}\label{eqn:extSC}
  {D^A D_A} \, ( f \cdot b ) = 0 \quad \forall\;\text{fluctuations}\;f\,,\,\text{background fields}\,b
\end{equation}
which connects the fluctuations with the background fields. As a consequence of this constraint, DFT$_{\mathrm{WZW}}$ reduces to DFT. More specifically it means that the two generalized Lie derivatives
\begin{equation}
  \mathcal{L}_\xi V^M = \mathcal{\hat{L}}_\xi V^M
\end{equation}
match under the extended strong constraint. In a similar fashion the actions
\begin{equation}
  S_{DFT_\mathrm{WZW}} = S_{DFT}
    \quad \text{and structure coefficients} \quad 
  \mathcal{F}_{\hat{A}\hat{B}\hat{C},DFT_{\mathrm{WZW}}} = \mathcal{F}_{\hat{A}\hat{B}\hat{C},DFT}
\end{equation}
are identical after imposing \eqref{eqn:extSC}.

It is important to note that this constraint is entirely optional and does not have to be imposed. As a consequence, it is reasonable to suspect that DFT$_\mathrm{WZW}$ can describe backgrounds which go beyond the scope of toroidal DFT. This even seems to be the case for geometric backgrounds, e.g. $S^3$ with $H$-flux \cite{Hassler:2016srl}, and not only for non-geometric backgrounds. Under the imposition of the extended strong constraint DFT$_\mathrm{WZW}$ becomes background independent. This result confirms the observations made in \cite{Hohm:2010jy}. Furthermore, employing this additional constraint rules out any solutions going beyond the SUGRA regime. Therefore, DFT$_\mathrm{WZW}$ possesses the same background independence as toroidal DFT but allows access to physics not attainable from SUGRA. Moreover, the constraint breaks the $2D$-diffeomorphism invariance of DFT$_\mathrm{WZW}$.

\section{Conclusion and Outlook}\label{sec:conclusion}
Starting from a purely geometric setup which gives rise to a WZW model with two equivalent Ka$\check{\text{c}}$-Moody algebras for the left and right moving parts of the closed string, one can derive Double Field Theory on Group Manifolds by using Closed String Field Theory. The effective action and gauge transformations are subsequently computed up to cubic order in a large level $k$ limit.

Afterwards, we have seen that Double Field Theory on Group Manifolds can be cast into a Generalized Metric Formulation and Flux Formulation very similar to the formulations known from toroidal Double Field Theory. As it turns out, this theory is invariant under generalized diffeomorphisms and double Lorentz transformations as well. We also saw that the background fields in our theory only need to fulfill the weaker Jacobi identity as opposed to the fluctuation fields which are always required to satisfy the strong constraint. These features are in stark contrast to toroidal DFT where all fields are constrained to obey the strong constraint. Furthermore, due to its underlying geometric structure, the group manifolds, it possesses an additional $2D$-diffeomorphism invariance which cannot be seen in toroidal DFT. This invariance breaks down during the transition from DFT$_\mathrm{WZW}$ to DFT, i.e. by imposing the extended strong constraint. It connects the background with the fluctuations of our theory. Once the extended strong constraint is imposed, DFT$_\mathrm{WZW}$ reduces to DFT. In this context, DFT$_\mathrm{WZW}$ generalizes the original DFT description and might be able to describe backgrounds inaccessible from DFT and SUGRA. It could even be that DFT$_\mathrm{WZW}$, due to its framework, allows for truly non-geometric backgrounds containing new physical information. Moreover, DFT$_\mathrm{WZW}$ enables us to construct an appropriate twist for each embedding tensor solution. A reason for this lies in the availability of all tools known from Riemannian geometry e.g. Maurer-Cartan forms which are unavailable in the toroidal DFT description. Additionally, an appropriate generalized Scherk-Schwarz compactification ansatz allowed us to recover half-maximal, electrically gauged SUGRA from DFT$_\mathrm{WZW}$.

The next step in our research will be the extension of the DFT$_{\mathrm{WZW}}$ framework to Exceptional Field Theories \cite{Bosque:2016fpi}. In this context, we are also going to answer the problem of solving the strong constraint/section condition in DFT$_\mathrm{WZW}$, see \cite{Hassler:2016srl}, and curved EFT. Especially, non-trivial solutions of the strong constraint/section condition which differ significantly from the original DFT/EFT formulation could reveal the full power of our framework. Furthermore, they could lead to a better insight into the functional principles of dualities, e.g. studying T-Duality in the case of $S^3$ as group manifold.
Another interesting question regards the possibility of extending the DFT/EFT setup to arbitrary background geometries, e.g. coset constructions. On top of that, DFT$_\mathrm{WZW}$ could be a great tool to analyze non-associativity and non-commutativity of non-geometric backgrounds. Last but not least, it would definitely be interesting to take higher $\alpha'$ corrections into account.


\begin{thebibliography}{99}

\bibitem{Tseytlin:1990va} 
  A.~A.~Tseytlin,
  ``Duality symmetric closed string theory and interacting chiral scalars,''
  Nucl.\ Phys.\ B {\bf 350}, 395 (1991).

\bibitem{Siegel:1993th}
  W.~Siegel,
  ``Superspace duality in low-energy superstrings,''
  Phys.\ Rev.\ D {\bf 48} (1993) 2826
  [hep-th/9305073].

\bibitem{Hull:2009mi}
  C.~Hull and B.~Zwiebach,
  ``Double Field Theory,''
  JHEP {\bf 0909} (2009) 099
  [arXiv:0904.4664 [hep-th]].

\bibitem{Hull:2009zb}
  C.~Hull and B.~Zwiebach,
  ``The Gauge algebra of double field theory and Courant brackets,''
  JHEP {\bf 0909} (2009) 090
  [arXiv:0908.1792 [hep-th]].
  
\bibitem{Hohm:2010pp}
  O.~Hohm, C.~Hull and B.~Zwiebach,
  ``Generalized metric formulation of double field theory,''
  JHEP {\bf 1008} (2010) 008
  [arXiv:1006.4823 [hep-th]].
  

\bibitem{Hohm:2010jy}
  O.~Hohm, C.~Hull and B.~Zwiebach,
  ``Background independent action for double field theory,''
  JHEP {\bf 1007} (2010) 016
  [arXiv:1003.5027 [hep-th]].

\bibitem{Hohm:2011ex}
  O.~Hohm and S.~K.~Kwak,
  ``Double Field Theory Formulation of Heterotic Strings,''
  JHEP {\bf 1106} (2011) 096
  [arXiv:1103.2136 [hep-th]].

\bibitem{Aldazabal:2013sca}
  G.~Aldazabal, D.~Marques and C.~Nunez,
  ``Double Field Theory: A Pedagogical Review,''
  Class.\ Quant.\ Grav.\  {\bf 30} (2013) 163001
  [arXiv:1305.1907 [hep-th]].

\bibitem{Hohm:2013bwa}
  O.~Hohm, D.~Lüst and B.~Zwiebach,
  ``The Spacetime of Double Field Theory: Review, Remarks, and Outlook,''
  Fortsch.\ Phys.\  {\bf 61} (2013) 926
  [arXiv:1309.2977 [hep-th]].

\bibitem{Buscher:1987sk}
  T.~H.~Buscher,
  ``A Symmetry of the String Background Field Equations,''
  Phys.\ Lett.\ B {\bf 194} (1987) 59.

\bibitem{Tseytlin:1990nb}
  A.~A.~Tseytlin,
  ``Duality Symmetric Formulation of String World Sheet Dynamics,''
  Phys.\ Lett.\ B {\bf 242} (1990) 163.
  
  \bibitem{Dabholkar:2002sy}
  A.~Dabholkar and C.~Hull,
  ``Duality twists, orbifolds, and fluxes,''
  JHEP {\bf 0309} (2003) 054
  [hep-th/0210209].

\bibitem{Dabholkar:2005ve}
  A.~Dabholkar and C.~Hull,
  ``Generalised T-duality and non-geometric backgrounds,''
  JHEP {\bf 0605} (2006) 009
  [hep-th/0512005].

\bibitem{Hull:2006va}
  C.~M.~Hull,
  ``Doubled Geometry and T-Folds,''
  JHEP {\bf 0707} (2007) 080
  [hep-th/0605149].

  
  \bibitem{Hull:2004in}
  C.~M.~Hull,
  ``A Geometry for non-geometric string backgrounds,''
  JHEP {\bf 0510} (2005) 065
  [hep-th/0406102].

  
  \bibitem{Hull:2009sg}
  C.~M.~Hull and R.~A.~Reid-Edwards,
  ``Non-geometric backgrounds, doubled geometry and generalised T-duality,''
  JHEP {\bf 0909} (2009) 014
  [arXiv:0902.4032 [hep-th]].
  
  \bibitem{Andriot:2012wx} 
  D.~Andriot, O.~Hohm, M.~Larfors, D.~Lüst and P.~Patalong,
  ``A geometric action for non-geometric fluxes,''
  Phys.\ Rev.\ Lett.\  {\bf 108}, 261602 (2012)
  [arXiv:1202.3060 [hep-th]].
  
  \bibitem{Hull:2005hk}
  C.~M.~Hull and R.~A.~Reid-Edwards,
  ``Flux compactifications of string theory on twisted tori,''
  Fortsch.\ Phys.\  {\bf 57} (2009) 862
  [hep-th/0503114].

\bibitem{Aldazabal:2011nj}
  G.~Aldazabal, W.~Baron, D.~Marques and C.~Nunez,
  ``The effective action of Double Field Theory,''
  JHEP {\bf 1111} (2011) 052
  [arXiv:1109.0290 [hep-th]].

\bibitem{Grana:2012rr}
  M.~Grana and D.~Marques,
  ``Gauged Double Field Theory,''
  JHEP {\bf 1204} (2012) 020
  [arXiv:1201.2924 [hep-th]].

\bibitem{Aldazabal:2013mya}
  G.~Aldazabal, M.~Graña, D.~Marqués and J.~A.~Rosabal,
  ``Extended geometry and gauged maximal supergravity,''
  JHEP {\bf 1306} (2013) 046
  [arXiv:1302.5419 [hep-th]].

\bibitem{Berman:2013cli}
  D.~S.~Berman and K.~Lee,
  ``Supersymmetry for Gauged Double Field Theory and Generalised Scherk-Schwarz Reductions,''
  Nucl.\ Phys.\ B {\bf 881} (2014) 369
  [arXiv:1305.2747 [hep-th]].

\bibitem{Hassler:2014sba}
  F.~Hassler and D.~Lüst,
  ``Consistent Compactification of Double Field Theory on Non-geometric Flux Backgrounds,''
  JHEP {\bf 1405} (2014) 085
  [arXiv:1401.5068 [hep-th]].
  
  \bibitem{Condeescu:2012sp}
  C.~Condeescu, I.~Florakis and D.~Lüst,
  ``Asymmetric Orbifolds, Non-Geometric Fluxes and Non-Commutativity in Closed String Theory,''
  JHEP {\bf 1204} (2012) 121
  [arXiv:1202.6366 [hep-th]].

\bibitem{Condeescu:2013yma}
  C.~Condeescu, I.~Florakis, C.~Kounnas and D.~Lüst,
  ``Gauged supergravities and non-geometric Q/R-fluxes from asymmetric orbifold CFT`s,''
  JHEP {\bf 1310} (2013) 057
  [arXiv:1307.0999 [hep-th]].
  
\bibitem{Blumenhagen:2014gva}
  R.~Blumenhagen, F.~Hassler and D.~Lüst,
  ``Double Field Theory on Group Manifolds,''
  JHEP {\bf 1502} (2015) 001
  [arXiv:1410.6374 [hep-th]].
  
  \bibitem{Blumenhagen:2015zma}
  R.~Blumenhagen, P.~du Bosque, F.~Hassler and D.~Lüst,
  ``Generalized Metric Formulation of Double Field Theory on Group Manifolds,''
  JHEP {\bf 1508} (2015) 056
  [arXiv:1502.02428 [hep-th]].

\bibitem{Bosque:2015jda}
  P.~du Bosque, F.~Hassler and D.~Lüst,
  ``Flux Formulation of DFT on Group Manifolds and Generalized Scherk-Schwarz Compactifications,''
  JHEP {\bf 1602} (2016) 039
  [arXiv:1509.04176 [hep-th]].
  
\bibitem{Hohm:2015ugy} 
  O.~Hohm and D.~Marques,
  ``Perturbative Double Field Theory on General Backgrounds,''
  Phys.\ Rev.\ D {\bf 93}, no. 2, 025032 (2016)
  [arXiv:1512.02658 [hep-th]].
  
  
\bibitem{Geissbuhler:2011mx}
  D.~Geissbuhler,
  ``Double Field Theory and N=4 Gauged Supergravity,''
  JHEP {\bf 1111} (2011) 116
  [arXiv:1109.4280 [hep-th]].

\bibitem{Hohm:2013nja}
  O.~Hohm and H.~Samtleben,
  ``Gauge theory of Kaluza-Klein and winding modes,''
  Phys.\ Rev.\ D {\bf 88} (2013) 085005
  [arXiv:1307.0039 [hep-th]].

\bibitem{Geissbuhler:2013uka}
  D.~Geissbuhler, D.~Marques, C.~Nunez and V.~Penas,
  ``Exploring Double Field Theory,''
  JHEP {\bf 1306} (2013) 101
  [arXiv:1304.1472 [hep-th]].

\bibitem{Hassler:2016srl} 
  F.~Hassler,
  ``The Topology of Double Field Theory,''
  arXiv:1611.07978 [hep-th].

\bibitem{Bosque:2016fpi}
  P.~du Bosque, F.~Hassler, D.~Lüst and E.~Malek,
  ``A geometric formulation of exceptional field theory,''
  arXiv:1605.00385 [hep-th].



\end{thebibliography}
\end{document}